\input harvmac.tex
%
\font\tenams=msam10 \font\sevenams=msam7 
\newfam\amsfam 
\textfont\amsfam=\tenams
\scriptfont\amsfam=\sevenams  

\def\hexnumber#1{\ifcase#1 0\or 1\or 2\or 3\or 4\or 5\or 6\or 7\or
8\or 9\or A\or B\or C\or D\or E\or F\fi}

\edef\theamsfam{\hexnumber\amsfam}  

\mathchardef\gtrsim="3\theamsfam 26 
\mathchardef\lesssim="3\theamsfam 2E 
\mathchardef\square"0\theamsfam 03

\ifx\epsfbox\UnDeFiNeD\message{(NO epsf.tex, FIGURES WILL BE
IGNORED)}
\def\figin#1{\vskip2in}
\else\message{(FIGURES WILL BE INCLUDED)}\def\figin#1{#1}\fi
\def\ifig#1#2#3{\xdef#1{fig.~\the\figno}
\midinsert{\centerline{\figin{#3}}%
\smallskip\centerline{\vbox{\baselineskip12pt
\advance\hsize by -1truein\noindent{\bf Fig.~\the\figno:} #2}}
\bigskip}\endinsert\global\advance\figno by1}
\noblackbox

\def\frac#1#2{{{#1} \over {#2}}}

%
%

%
\lref\shenker{
N.~Kaloper, M.~Kleban, A.~E.~Lawrence and S.~Shenker,
``Signatures of short distance physics in the cosmic microwave  background,''
arXiv:hep-th/0201158.
}
\lref\StarobinskyRP{
A.~A.~Starobinsky and I.~I.~Tkachev,
``Trans-Planckian particle creation in cosmology and ultra-high energy  cosmic rays,''
arXiv:astro-ph/0207572.
}
\lref\danielssontwo{
U.~H.~Danielsson,
``Inflation, holography and the choice of vacuum in de Sitter space,''
JHEP {\bf 0207}, 040 (2002)
[arXiv:hep-th/0205227].
}
\lref\EastherXE{
R.~Easther, B.~R.~Greene, W.~H.~Kinney and G.~Shiu,
``A generic estimate of trans-Planckian modifications to the primordial  power spectrum in inflation,''
Phys.\ Rev.\ D {\bf 66}, 023518 (2002)
[arXiv:hep-th/0204129].
}
\lref\danielsson{
U.~H.~Danielsson,
``A note on inflation and transplanckian physics,''
Phys.\ Rev.\ D {\bf 66}, 023511 (2002)
[arXiv:hep-th/0203198].
}
\lref\BrandenbergerNQ{
R.~Brandenberger and P.~M.~Ho,
``Noncommutative spacetime, stringy spacetime uncertainty principle, and  density fluctuations,''
Phys.\ Rev.\ D {\bf 66}, 023517 (2002)
[arXiv:hep-th/0203119].
}
\lref\LizziIB{
F.~Lizzi, G.~Mangano, G.~Miele and M.~Peloso,
``Cosmological perturbations and short distance physics from  noncommutative geometry,''
JHEP {\bf 0206}, 049 (2002)
[arXiv:hep-th/0203099].
}
\lref\NiemeyerKH{
J.~C.~Niemeyer, R.~Parentani and D.~Campo,
``Minimal modifications of the primordial power spectrum from an  adiabatic short distance cutoff,''
arXiv:hep-th/0206149.
}
\lref\LoweAC{
D.~A.~Lowe, J.~Polchinski, L.~Susskind, L.~Thorlacius and J.~Uglum,
``Black hole complementarity versus locality,''
Phys.\ Rev.\ D {\bf 52}, 6997 (1995)
[arXiv:hep-th/9506138].
}
\lref\KempfFA{
A.~Kempf and J.~C.~Niemeyer,
``Perturbation spectrum in inflation with cutoff,''
Phys.\ Rev.\ D {\bf 64}, 103501 (2001)
[arXiv:astro-ph/0103225].
}
\lref\BrandenbergerHS{
R.~H.~Brandenberger and J.~Martin,
``On signatures of short distance physics in the cosmic microwave  background,''
arXiv:hep-th/0202142.
}
\lref\FloreaniniTQ{
R.~Floreanini, C.~T.~Hill and R.~Jackiw,
``Functional Representation For The Isometries Of De Sitter Space,''
Annals Phys.\  {\bf 175}, 345 (1987).
}
\lref\EastherFZ{
R.~Easther, B.~R.~Greene, W.~H.~Kinney and G.~Shiu,
``Imprints of short distance physics on inflationary cosmology,''
arXiv:hep-th/0110226.
}
\lref\EastherFI{
R.~Easther, B.~R.~Greene, W.~H.~Kinney and G.~Shiu,
``Inflation as a probe of short distance physics,''
Phys.\ Rev.\ D {\bf 64}, 103502 (2001)
[arXiv:hep-th/0104102].
}
\lref\MartinKT{
J.~Martin and R.~H.~Brandenberger,
``The Corley-Jacobson dispersion relation and trans-Planckian inflation,''
Phys.\ Rev.\ D {\bf 65}, 103514 (2002)
[arXiv:hep-th/0201189].
}
\lref\HuiCE{
L.~Hui and W.~H.~Kinney,
``Short distance physics and the consistency relation for scalar and  tensor fluctuations in the inflationary universe,''
Phys.\ Rev.\ D {\bf 65}, 103507 (2002)
[arXiv:astro-ph/0109107].
}
\lref\allen{
B.~Allen,
``Vacuum States In De Sitter Space,''
Phys.\ Rev.\ D {\bf 32}, 3136 (1985).
}
\lref\mottola{
E.~Mottola,
``Particle Creation In De Sitter Space,''
Phys.\ Rev.\ D {\bf 31}, 754 (1985).
}
\lref\MartinXS{
J.~Martin and R.~H.~Brandenberger,
``The trans-Planckian problem of inflationary cosmology,''
Phys.\ Rev.\ D {\bf 63}, 123501 (2001)
[arXiv:hep-th/0005209].
}
\lref\BrandenbergerWR{
R.~H.~Brandenberger and J.~Martin,
``The robustness of inflation to changes in super-Planck-scale physics,''
Mod.\ Phys.\ Lett.\ A {\bf 16}, 999 (2001)
[arXiv:astro-ph/0005432].
}
\lref\BrandenbergerSW{
R.~H.~Brandenberger,
``Inflationary cosmology: Progress and problems,''
arXiv:hep-ph/9910410.
}
\lref\StarobinskyKN{
A.~A.~Starobinsky,
``Robustness of the inflationary perturbation spectrum to trans-Planckian  physics,''
Pisma Zh.\ Eksp.\ Teor.\ Fiz.\  {\bf 73}, 415 (2001)
[JETP Lett.\  {\bf 73}, 371 (2001)]
[arXiv:astro-ph/0104043].
}

\Title{ \vbox{\baselineskip12pt\hbox{hep-th/0208167}
\hbox{BROWN-HET-1329}
}}
{\vbox{
\centerline{Initial state effects on the cosmic microwave}
\centerline{ background
and trans-planckian physics}}}

\centerline{Kevin Goldstein and David A. Lowe}
\medskip

\centerline{Department of Physics}
\centerline{Brown University}
\centerline{Providence, RI 02912, USA}
\centerline{\tt kevin, lowe@het.brown.edu}
\bigskip

\centerline{\bf{Abstract}}
There exist a one complex parameter family of de Sitter invariant
vacua, known as $\alpha$ vacua. 
In the context of slow roll inflation, we show that all but the
Bunch-Davies vacuum generates unacceptable production of high
energy particles
at the end of inflation. As a simple model for the effects of
trans-planckian physics, we go on to consider non-de
Sitter invariant vacua obtained by patching modes in the Bunch-Davies
vacuum above some momentum scale $M_c$, with modes in an $\alpha$
vacuum below $M_c$. Choosing $M_c$ near
the Planck scale $M_{pl}$,
we find acceptable levels of hard particle production, and corrections to the
cosmic microwave perturbations at the level of $H M_{pl}/M_c^2$, where $H$
is the Hubble parameter during inflation.
More general initial states of this type  with $H\ll M_c \ll M_{pl}$ can
give corrections to the spectrum of cosmic microwave background
perturbations at order 1. The parameter characterizing the
$\alpha$-vacuum during inflation is a new cosmological observable.

\vfill
\Date{\vbox{\hbox{\sl August, 2002}}}

\newsec{Introduction}

Inflation magnifies quantum fluctuations at fundamental length scales
to astrophysical scales, where their imprint is left on the formation
of structure in the universe. In conventional slow roll inflation,
the universe undergoes an expansion of at least $10^{26}$ during the
inflationary phase. With such a huge expansion factor, modes which
give rise to observable structures apparently started out with
wavelengths much smaller than the Planck length. This is the so-called
trans-planckian problem in inflation
\refs{\BrandenbergerSW \BrandenbergerWR \MartinXS \MartinKT {--}\StarobinskyKN}. 

In the past year, there has been much debate about whether potential
modifications to physics above the Planck scale could actually be
observed \refs{\EastherFI \EastherFZ \BrandenbergerHS
\BrandenbergerNQ \LizziIB \danielsson \EastherXE \danielssontwo
\shenker \NiemeyerKH \KempfFA {--}\StarobinskyRP}.  By considering the local effective action
at the Hubble scale $H$ (which we will take to be
$10^{13}-10^{14}$ GeV), \shenker\ has argued that trans-planckian
corrections to the spectrum of cosmic microwave background
perturbations could at best be of order $(H/M_{pl})^2$ which is
typically far too small to be observed in conventional inflationary
models. However others \refs{\EastherFI \EastherFZ \BrandenbergerHS
\BrandenbergerNQ \danielsson \EastherXE {--}\danielssontwo} have obtained
a correction of order $H/M_{pl}$ by considering a variety of methods
for modeling trans-planckian effects. Such a correction is potentially observable
in the not too distant future.

In the present work we represent the effect of trans-planckian physics
simply by allowing for nontrivial initial vacuum states for the
inflaton field, which we treat as a free scalar field moving in a de
Sitter background. The most natural vacuum states to consider are the
de Sitter invariant vacuum states constructed by Allen and Mottola
\refs{\allen,\mottola}. The vacuum states are known as $\alpha$-vacua.
We find these all lead to infinite energy
production at the end of inflation, with the exception of the
Bunch-Davies (Euclidean) vacuum state.

We go on to consider non-de Sitter invariant vacuum states constructed
by placing modes with comoving wavenumber $k > M_c a(\eta_f)$ in the
Bunch-Davies vacuum, where
$a(\eta_f)$ is the expansion factor at the end of inflation. Modes
with $k < M_c a(\eta_f)$ are placed in a non-trivial $\alpha$ vacuum.
These states have a particularly simple evolution in de Sitter space
-- the length scale at which the patching occurs simply expands as the scale
factor grows. Many more complicated initial states asymptote to such
states as the universe expands.

For $M_c$ of order $M_{pl}$ it is possible to find initial states that
do not overproduce hard
particles, and produce corrections to the cosmic microwave background
spectrum at order $H/M_{c}$ in agreement with \EastherFZ. For $H\ll M_c
\ll M_{pl}$  there are initial states that produce corrections to the
 spectrum at order 1. 

In \refs{\danielsson,\danielssontwo}
an initial state was constructed by placing modes in
their locally Minkowskian 
vacuum states as the proper wavenumber passed through the scale
of new physics $M_c$. This turns out to be a special case of the class
of initial states we consider. To avoid large backreaction problems in
this case, we show the condition $M_c \ll M_{pl}$ must hold. This condition is
rather easy to satisfy. Our more general initial states may be viewed
in a similar way
as an initial state that puts modes in a $k$-independent Bogoliubov
transformation of the locally Minkowskian vacuum as proper wavenumber
passes through the scale $M_c$.

\newsec{General setup}

We will conduct our analysis using linearized perturbation theory in a
de Sitter (dS) background.
Planar coordinates covering half of dS, with flat spacial sections, 
result in the metric 
\eqn\metone{ 
ds^2 = dt^2 - e^{2Ht}d\vec{x}^2 = dt^2 - a^2(t)d\vec{x}^2~.
}
It will be more convenient to use conformal coordinates, giving
\eqn\mettwo{
  ds^2 = \frac{1}{(\eta H)^2}\left( d\eta^2 - d\vec x^2 \right) 
  = a^2(\eta)\left( d\eta^2 - d\vec x^2 \right)
}
where $\eta = \int^\infty_t dt'/a(t')= -\exp(-Ht)/H$. So 
$t \to -\infty$ and $\eta \to -\infty$, and $t\to \infty$ as $\eta \to
0$.

Klein-Gordon Equation in curved space is
\eqn\klein{
 (\square + m^2+\zeta R)\phi=0
}
for a scalar field with mass $m$ and non-minimal coupling to $R$ given
by $\zeta$. In momentum space we can solve this equation by defining
\eqn\ansatz{
 \phi_k = 
  \frac{e^{i \vec k \cdot \vec x}}{(2\pi)^{\frac{3}{2}} a(\eta)}
  \chi_k(\eta) 
}
which leads to 
\eqn\eom{
 \chi_k'' + \left(k^2 + \frac{M^2}{H^2\eta^2}\right)\chi_k =0
}
with 
\eqn\Meqn{
 M^2= m^2 + (\zeta-\frac{1}{6})R
}
so $M^2$ is not necessarily positive.  The general solution is
\eqn\eomsol{
 \chi_k(\eta)=\half \sqrt{\pi\eta}~H^{(2)}_{\nu}(k \eta) \equiv \chi_{Ek}(\eta)
}
together with its complex conjugate, 
where $\nu=\frac{9}{4}-\frac{m^2}{H^2}-12\zeta= \frac{1}{4}-M^2$. 

Such a complete set of orthonormal modes may be used to define a Fock
vacuum state by taking the field operator
\eqn\phiop{
\hat \chi = \sum_k \chi_k(\eta) a_k + \chi_k^*(\eta) a^\dagger_{-k}
}
and demanding $a_k | 0 \rangle =0$.
As
shown by Allen \allen\ and Mottola \mottola, 
the general family of de Sitter invariant
vacuum states can be defined using the modes
\eqn\avacs{
\chi_k = \cosh \alpha ~\chi_{Ek}(\eta) +
e^{i \delta} \sinh\alpha ~\chi_{Ek}^*(\eta)
}
with  $\alpha \in  [0,\infty)$ and $\delta\in (-\pi,\pi)$.
$\alpha=0$ is the Bunch-Davies vacuum (a.k.a. Euclidean
vacuum). 

For a massless minimally coupled scalar, this solution takes a
particularly simple form
\eqn\massless{
\chi_{Ek} (\eta) = \frac{e^{-i k \eta}}{\sqrt{2k}} (1 - \frac{i}{k \eta})~.
}
As discussed in \allen\ this case gives rise to
difficulties in canonical quantization, and there is no de Sitter
invariant Fock vacuum. Nevertheless, we will use this simple example
in the following with the understanding a small mass term could be
added to eliminate this problem, and the expressions we
derive will not be substantially changed.

We will need to extract two physical quantities from the expression
\avacs. The first is the number of particles produced in the mode $k$
defined with respect to the $\alpha=0$ vacuum. This is simply equal to
\eqn\modeprod{
n_k = \sinh^2 \alpha~.
}
This will be a good
approximation to the number of particles produced at the end of
inflation, when a transition is made to a much more slowly expanding
universe, provided the wavelength of the modes in question are much
smaller than the Hubble radius. This follows simply from the fact that
at high wavenumber, the wave equation for $\chi_k$ reduces to that of
flat space, so we can approximate the final geometry by Minkowski
space. We wish to count particles with respect to the Lorentz
invariant vacuum state, which corresponds to the $\alpha=0$ vacuum in
this regime. 
  
The second physical quantity of interest 
is the contribution of this mode to the spectrum of CMBR
perturbations. We compute this by examining $|\phi_k(\eta)|^2$ 
in the distant future $\eta \to 0$ for the
massless scalar \massless.  
The contribution is then
\eqn\cmbpert{
P_k = \frac{k^3}{2\pi^2 a^2} |\chi_k|^2=\left(\frac{H}{2\pi}\right)^2 |\cosh \alpha - e^{i\delta} \sinh \alpha |^2~.
}

\newsec{Initial state effects}

We begin by reviewing what happens for the usual Bunch-Davies vacuum,
$\alpha=0$. Clearly the particle production at high frequencies
\modeprod\ vanishes. 
Fluctuations in the scalar field modes mean
different regions of spacetime expand at slightly different rates,
which gives rise to density perturbations after inflation has
ended. The amplitude of these perturbations are frozen in as these
modes expand outside the Hubble radius during inflation, and become
density perturbations once they reenter the horizon after the end of
inflation. For $\alpha=0$, $P_k = (H/2\pi)^2$ is independent of $k$
and hence scale invariant. When one allows for the detailed shape of
the inflaton potential, $H$ becomes effectively $k$ dependent, leading
to small deviations from the scale invariant spectrum of
perturbations, which in general are highly model dependent.

For a nontrivial $\alpha \neq 0$ vacuum we immediately see a
problem. At the end of inflation there will be a large amount of
particle production at wavelengths smaller than the Hubble radius
\modeprod. Since this production is independent of $k$ \allen, this will lead
to an infinite energy density, and singular backreaction on the geometry. 
We conclude then that at wavelengths below some scale, the modes must
be in a local $\alpha=0$ vacuum state.\foot{\FloreaniniTQ\ also
concludes that only the Euclidean vacuum smoothly patches onto the
Lorentz invariant Minkowski vacuum, in the context of two-dimensional
de Sitter space. They also point out that for all $\alpha \neq 0$ the
vacuum state picks up a nontrivial phase under de Sitter isometries,
which cancels in expectation values.} Actually, $\alpha$ need not be
exactly zero for the high wavenumber modes. We will return to this
point at the end of this section.

Nevertheless, we can still consider initial states that involve modes
in an $\alpha \neq 0$ state, provided their wavelengths are
sufficiently large. Perhaps the simplest such initial state is to
place modes at some fixed conformal time $\eta_0$ in the $\alpha=0$
state for $k> M_c a(\eta_f)$ where $\eta_f$ is the conformal time at
the end of inflation, and $M_c$ is some scale at which physics
changes, and we have in mind taking $M_c \gg H$.
Modes for $k< M_c a(\eta_f)$ can be placed in an $\alpha \neq 0$
state.

In order that the particle production at the end of inflation be
irrelevant versus the energy stored in the inflaton, we must have
\eqn\nobadpart{
M_c^4 \sinh^2 \alpha \ll \Lambda= \frac{3 M_{pl}^2 H^2}{4\pi}
}
where $M_{pl}$ is the Planck mass.\foot{This condition is necessary to
avoid large back-reaction on the geometry. It would also be
interesting to consider the limit when this energy is not
irrelevant, and to use this particle production as a source for
reheating.} 
If we saturate this bound, $\sinh \alpha \sim H M_{pl}/M_c^2$. The
correction to the  CMBR spectrum $P_k$ \cmbpert\ will then be of order
$H M_{pl}/M_c^2$. This is linear in $H$ in agreement  with the
estimates of 
\refs{\EastherFI,\EastherFZ,\danielsson}
 and is potentially observable. Of course, since we have
done the computation in pure de Sitter space, the effect appears as a
$k$-independent modulation of the $\alpha=0$ result, which on its own
would require an independent determination of $H$ to measure directly. However, in 
inflation $H$ is actually slowly changing, which will translate into
$k$-dependence of $H$, and hence $\alpha$. This will show up as
$k$-dependent corrections to the cosmic microwave background spectrum
$P_k$ which are potentially more easily distinguishable from the
$\alpha=0$ case \refs{\EastherFI,\EastherXE}.

To obtain an upper bound on the size of the correction to the CMBR spectrum,
we can imagine taking $M_c$ to be much smaller than $M_{pl}$, which is
certainly plausible. This allows $\alpha$ to be of order 1, and still
consistent with negligible hard particle production \nobadpart.
This limit will lead to
corrections to the CMBR spectrum \cmbpert\ at order 1.

\subsec{Transition at proper energy $M_c$}

Now let us consider a more detailed model for the initial state where
we assume the initial condition is fixed due to some change in physics
at the proper energy scale $M_c$.
Let us review the computation of \refs{\danielsson,\danielssontwo}. 
The essential idea was to note the $\alpha$-vacuum satisfying
\eqn\danrev{
\eqalign{
\cosh \alpha &= e^{i(\gamma- \beta)} \frac{2\beta -i}{2 \beta}\cr
e^{i\delta} \sinh \alpha & = - e^{i (\gamma+\beta)} \frac{i}{2\beta}
}}
with $\beta= M_c/H$ and $\gamma$ real, can be interpreted as an initial state which
places modes in their locally Minkowskian vacuum as the proper
wavenumber $k/a$ passes through the scale $M_c$. This is seen by
noting that at time $\eta = -M_c/H k$ the field $\phi_k$ (with
$\chi_k$ given by \avacs)
satisfies $\pi_k = -i k \phi_k$ where $\pi_k$ is the conjugate
momentum. Such a relation is satisfied by the Lorentz invariant vacuum
in Minkowski space. One may also interpret the state at time
$\eta=-M_c/H k$ as a minimum uncertainty state \danielsson.

For sufficiently large $k$, the above
prescription does not apply, because the relevant time $\eta$
will be after the end of inflation. These modes may safely be placed
in the Bunch-Davies vacuum.

This initial state is a special case of the type described above. 
High frequency particle creation at the end of inflation 
gives an energy density of order $M_c^4 \sinh^2 \alpha$. 
Since here $\sinh \alpha \sim H/M_c$, we require 
\eqn\niceobs{
M_c^2 H^2
\ll M_{pl}^2 H^2~.
}
This will hold whenever $M_c \ll M_{pl}$,
which is easy to satisfy. This condition was also obtained in \EastherFI.

Note that the general class of initial states described above may be
reinterpreted in the same way as states arising from a boundary condition placed at a
fixed proper energy scale. Rather than imposing the condition that the
initial state corresponds to a locally Minkowski vacuum as the
wavenumber $k/a$ passes through $M_c$, one instead demands the mode be
in a general $k$-independent Bogoliubov transformation of the locally
Minkowski vacuum. This corresponds to a generic boundary condition at
the scale $M_c$ that is independent of time.
In this way, modes are placed in a nontrivial
$\alpha$-vacuum when $k/a(\eta_f)$ at the end of inflation is
below the scale $M_c$. Higher $k$ modes will remain in the
Bunch-Davies vacuum. This is precisely the type of state we described
above.

It is interesting to view this boundary condition in the context of
the nice slice argument of \LoweAC\ used to define effective field
theory in a curved background. The conformal time slicing \mettwo\
satisfies the criteria for a ``nice slicing''. This mean we may define
fields with, for example, a spatial lattice cutoff on proper wavelengths below
$1/M_c$. As one moves forward on these time slices, the proper
wavelength of a given mode \eomsol\ expands, so new modes descend from
above the cutoff scale, and we assume these are placed in their ground
state. The main difference with asymptotically flat space, is that we
now have the option of placing these modes in one of the nontrivial de
Sitter invariant $\alpha$-vacua. Any other choice would lead to
continuous creation of particles at the cutoff scale which would cause
drastic back-reaction on the geometry. 

Provided interacting quantum field
theory in de Sitter is consistent in a general $\alpha$ vacuum, there seems to be
no dynamics that prefers one value of $\alpha$ over another. Only when
we patch de Sitter space onto standard cosmology at the end of
inflation do we generate observable consequences of the $\alpha$
parameter in the form of extra particle production, and imprint on the
CMBR. In the context of slow roll inflation, we should therefore
regard the value of $\alpha$ during inflation 
as a new cosmological observable which encodes
information about trans-planckian physics.

At the end of inflation we make a transition from the de Sitter
geometry to a standard cosmological geometry. To describe the UV
cutoff in this more general context, we need to replace the simple
$\alpha$-vacuum suitable for de Sitter, by a boundary condition fixed
by some more general dynamical condition
such as the locally Minkowskian boundary condition
of \refs{\danielsson, \danielssontwo} described above \danrev. The effective
value of $\alpha$ will then change as the effective value of $H$
changes. Note for us $H$ determines the vacuum energy density, and
is not related to the Hubble parameter outside the de Sitter phase.
In the limit that the cosmological constant becomes very
small (the effective $H$ decreases by a factor of $10^{-30}$ or so to match with
today's vacuum energy density), we make a smooth transition to a
$\alpha \sim 10^{-30} H/M_c$
boundary condition
at the
cutoff scale
after the end of inflation. If we regard the present state of the
universe as a de Sitter phase with very small cosmological constant,
this type of boundary condition does not lead to continuous particle
creation, so is not subject to the constraints explored in
\StarobinskyRP.

The value of $\alpha$ during inflation may be selected by 
local physics at Planckian energies, but in general $\alpha$ may
also be influenced by the initial state of the universe. This initial
state is not necessarily completely determined by physics at Planckian
energies. For example the initial state may emerge as a special state of very
high symmetry as a result of dynamics on much higher energy scales,
which will leave their imprint on the value of $\alpha$ in the de
Sitter phase.

\newsec{Conclusions}

We have constructed a very simple class of initial states for the
inflaton field which can be used to model effects of trans-planckian
physics. A new cosmological observable emerges from this analysis in
the context of slow-roll inflation, namely the $\alpha$ parameter
characterizing the vacuum state during inflation.

Other 
previous approaches have  typically assumed some definite
model for the trans-planckian physics which led to particular states of
this type at momenta much below the Planck scale.
We have found for certain ranges of parameters, the initial states do not lead to
excess
particle production at the end of inflation, and lead to potentially
observable corrections to the cosmic microwave background spectrum.

\bigskip
{\bf Acknowledgments}
We thank R. Brandenberger, R. Easther, B. Greene and B. Lehmann for helpful
discussions and comments.
D.L. thanks the
Aspen Center for Physics for hospitality during the completion of this
work, and the participants at the Center for stimulating discussions.
This research is supported in part by DOE
grant DE-FE0291ER40688-Task A and by BSF grant 2000359.

\listrefs

\end